%% file: DplusXsecPRL_v2.0.tex
\documentclass[prl,twocolumn,showpacs,superscriptaddress,floatfix]{revtex4-1}

\usepackage{graphics,overpic, rotating}  
\usepackage{makeidx}
\usepackage{colordvi}
\begin{document}  

\title{Measurement of the $D^+$-meson production cross section at low transverse momentum in $p\bar{p}$ collisions at $\sqrt{s}=1.96\,\mbox{TeV}$}

%% Author list here
\input{author.tex} 
\date{\today}
%% Abstract
\begin{abstract} 
We report on a measurement of the $D^{+}$-meson production cross section as a function of transverse momentum ($p_T$) in proton-antiproton ($p\bar{p}$) collisions at 1.96 TeV center-of-mass energy, using the full data set collected by the Collider Detector at Fermilab in Tevatron Run II and corresponding to 10 fb$^{-1}$ of integrated luminosity. We use $D^{+} \to K^-\pi^+\pi^+$ decays fully reconstructed in the central rapidity region $|y|<1$ with transverse momentum down to 1.5 GeV/$c$, a range previously unexplored in $p\bar{p}$ collisions. Inelastic $p\bar{p}$-scattering events are selected online using minimally-biasing requirements followed by an optimized offline selection. The $K^-\pi^+\pi^+$ mass distribution is used to identify the $D^+$ signal, and the $D^+$ transverse impact-parameter distribution is used to separate prompt production, occurring directly in the hard scattering process,
 from secondary production from $b$-hadron decays. We obtain a prompt $D^+$ signal of 2950 candidates corresponding to a total cross section $\sigma(D^+, 1.5 < p_T < 14.5~\mbox{GeV/}c, |y|<1) = 71.9  \pm 6.8 (\mbox{stat}) \pm 9.3 (\mbox{syst})~\mu$b. While the measured cross sections
are consistent with theoretical estimates in each $p_T$ bin, the shape of the observed $p_T$ spectrum is softer than the expectation from quantum chromodynamics. The results are unique in $p\bar{p}$ collisions and can improve the shape and uncertainties of future predictions.

%A two-dimensional simultaneous fit of the  $K^-\pi^+\pi^+$-mass and $D^{+}$ transverse-impact-parameter distributions yields a signal of 2950 $D^{+}$ candidates produced directly in the hard scattering process. The total cross section is measured to be $\sigma(D^+, 1.5 < p_T < 14.5~\mbox{GeV/}c, |y|<1) = 71.9  \pm 6.8 (\mbox{stat}) \pm 9.3 (\mbox{syst})~\mu$b. While the measured cross-section values as a function of $p_T$ lie within the band of theoretical uncertainty, differences in shape suggest that these results can improve the shape and uncertainties of the predictions.
%help further refinement of theoretical calculations. 
\end{abstract}

\maketitle    

%% Main text body
Measurements of cross sections for the production of hadrons containing bottom or charm quarks (heavy-flavors) in hadron collisions offer fundamental information to test and refine phenomenological models of the strong interaction at small momentum transfer, a regime in which perturbative expansions are challenging. In addition, in searches for astrophysical neutrinos, knowledge of charm production cross-sections may improve estimations of background rates from neutrinos produced in decays of charm hadrons from cosmic-ray interactions with atmospheric nuclei~\cite{charm-neutrinos}.\par
%During Run I (1992--1996) of the Fermilab Tevatron collider, the CDF experiment pioneered the study of bottom-meson cross-sections, using proton-antiproton ($p\bar{p}$) collisions at center-of-mass energy of $\sqrt s=1.8$\,TeV
The first studies of heavy-flavor production performed at the Tevatron proton-antiproton ($p\bar{p}$) collider in 1992--1996~\cite{RunI-BXSec} yielded cross sections significantly larger than the predicted values~\cite{QCDcalc} and prompted a dedicated effort in refining calculations~\cite{QCDcalc2}, which resulted in reduced discrepancies. The program continued during Tevatron Run II (2001--2011), including first measurements of charm-meson cross sections using $p\bar{p}$ collisions at center-of-mass energy $\sqrt{s} = 1.96$ TeV~\cite{Acosta:2003ax}. Since 2010, CERN's  LHC $pp$ collider has replaced the Tevatron as the most prolific charm-meson source, allowing the ALICE and LHCb experiments to report measurements of charm cross sections at $ \sqrt{s}$ = 2.76--13.00 TeV \cite{LHC}. \par Measurements based on $p\bar{p}$ collisions, and probing different collision energies, remain essential to extend the understanding of quantum chromodynamics (QCD), because differing admixtures of parton-level processes contribute at different energies and initial states. Previous measurements in $p\bar{p}$ collisions~\cite{Acosta:2003ax} were restricted to mesons with transverse momentum $p_T > 6.0$ GeV/$c$ because of the transverse-momentum thresholds used in the online event selection (trigger). The transverse momentum is the momentum component in the plane transverse to the beam.  Extending the reach to lower $p_T$, hence further into the nonperturbative regime, provides novel and unique constraints to improve QCD phenomenological models. \par In this paper, we report on a measurement of the production cross section for $D^+$ mesons down to 1.5 GeV/$c$ $p_T$, a range unexplored in $p \bar{p}$ collisions, and unlikely to be explored in the foreseeable future with this initial state. The measurement is performed as a function of meson transverse momentum using $D^{+} \to K^-\pi^+\pi^+$ decays reconstructed in the full CDF Run II data set, corresponding to 10~fb$^{-1}$ of integrated luminosity. Throughout this paper charge-conjugate decays are implied. Candidate $D^+$ signal events are selected from a minimum-bias sample, collected by imposing minimal requirements on the event features in order to minimize biases on the physics properties of charm decays.  
Events are divided into independent subsamples ($p_T$ bins) according to the $D^+$ candidate $p_T$. In each, we apply a data-driven optimization of the offline selection and perform a two-dimensional simultaneous fit of the resulting distributions of the $K^-\pi^+\pi^+$-mass and $D^{+}$ impact-parameter, defined as the minimum transverse distance between a particle's trajectory and the beam. The fit determines, for each $p_T$ bin, the prompt $D^+$ yield ($D^+$ mesons directly produced in the $p\bar{p}$ interaction or originating from charm resonances) by statistically subtracting secondary $D^+$ candidates ($D^+$ mesons originating from $b$-hadron decays).
%The price is a reduced fraction of events containing heavy mesons.  However, in the full 10 fb$^{-1}$ sample of Run II data, the size of such minimally-biased samples is sufficient to allow reconstruction of low-background samples charm-meson decays. 
Each prompt yield is combined with the corresponding reconstruction and selection efficiencies, derived using simulation, to determine the cross section,  
%The cross section $\sigma_i$, averaged in each $p_T$ bin $i$ and integrated in the range $|y|<1$  (where $y=0.5\ln[(E+p_z)/(E-p_z)]$ and $E$ and $p_z$ are charm-meson energy and longitudinal momentum, respectively) is determined as 
\begin{equation}
\sigma_i=\frac{N_i/2 }{\int\mathcal{L}dt\cdot\epsilon_i\cdot {\cal B}}~,
\end{equation}
where $N_i$ is the observed number of prompt $D^+$ and $D^-$ mesons in the $i$th $p_T$ bin. The factor $1/2$ is included because both $D^+$ and $D^-$ mesons contribute to $N_i$ and we report results solely for $D^+$, assuming charge-symmetric production of charm quarks in the strong $p\bar{p}$ interaction. 
The integrated luminosity $\int\mathcal{L}dt$ is normalized to an inelastic cross section of $\sigma_{p\bar{p}}=60.7\pm2.4$\,mb~\cite{inelastic} and $\epsilon_i$ is the global detection, reconstruction, and selection efficiency.  The branching fraction used for the $D^+ \to K^-\pi^+\pi^+$ decay is  ${\cal B} = (9.46 \pm 0.24)\%$~\cite{pdg}.\par
%The rate of inelastic colisions is measured with Cherenkov luminosity counters~\cite{Lerr} and has an uncertainty of 4.4\%.
%We verified that the $D$ and $\bar{D}$ cross sections are equal within statistical uncertainties.
%The differential production cross section $d\sigma/dp_T$ is intended to be integrated over $y$ between $-1$ and $+1$ and averaged over each $p_T$ bin.
%We report here measurements of prompt charm meson cross sections
%using data recorded with this trigger
%in February and March 2002, corresponding to $5.8\pm0.3\,\mbox{pb}^{-1}$ of integrated luminosity.
%%%%%%%%%%%%%%%%%
% Detector 
%%%%%%%%%%%%%%%%%
The CDF II detector is a multipurpose magnetic spectrometer surrounded by calorimeters and muon detectors~\cite{CDF}. It is roughly cylindrically symmetric around the beams and is described in a cylindrical coordinate system with the $z$ axis along the incident proton beam direction. The detector components relevant for this analysis are as follows. A silicon microstrip vertex detector and a cylindrical open-cell drift chamber immersed in a nearly uniform $1.4$~T axial magnetic field allow the reconstruction of charged-particle trajectories (tracks) in the pseudorapidity range $|\eta| < 1$. The vertex detector contains seven concentric layers of single- and double-sided silicon sensors at radii between 1.5 and 22~cm, each providing a position measurement with up to 15 (70)~$\mu$m resolution in the azimuthal (longitudinal) direction~\cite{Sill:zz}. The drift chamber has 96 measurement layers, located between 40 and 137~cm in radius, organized into alternating axial and $\pm 2^{\circ}$ stereo superlayers~\cite{COT}. The transverse momentum is determined with a resolution of $\sigma_{p_T}/p_T^2 \approx 0.07\%$ (GeV/$c$)$^{-1}$, corresponding to a typical mass resolution of 6.0~MeV/$c^2$ for a $D^+ \to K^-\pi^+\pi^+$ decay.  Gas Cherenkov detectors (CLC) covering the symmetric regions at small polar angle  around the interaction region $3.7 < |\eta| < 4.7$ are used to detect hard-scatter interactions and measure luminosity~\cite{lumi}.
%%%%%%%%%%%%%%%%%
% Trigger
%%%%%%%%%%%%%%%%%
CDF has a three-level trigger system. We use events collected by the zero- and minimum-bias triggers, which are designed to collect events while introducing minimal bias in the properties of the particles produced in the collision. The zero-bias trigger applies no selection requirements and accepts a $10^{-6}$ fraction (prescale factor) of $p\bar{p}$ crossings, randomly chosen. At the first trigger level, the minimum-bias trigger accepts a $10^{-5}$ prescale fraction of the events in which a time-coincidence between signals in the CLC at opposite sides of the interaction region is detected, which enriches the sample in $p\bar{p}$ crossings that yield inelastic interactions. At the second (third) trigger level, the minimum-bias trigger applies no requirements and accepts events with 3 (1) Hz maximum rates. The large prescale factors and accept-rate reductions avoid saturation of the data-writing rate. The resulting samples contain 183 million zero-bias and 133 million minimum-bias events. Of these, 409 events are common to both samples and used only once in the analysis.

%charged tracks are reconstructed in the COT axial projection
%by a hardware processor (XFT)~\cite{Thomson:2002xp}.
%The trigger for hadronic charm decays requires two oppositely
%charged tracks with $p_T\ge 2\,\mbox{GeV}/c$ and
%the scalar sum of the $p_T$'s larger than 5.5\,GeV/$c$,
%where $p_T$ is the magnitude of the component of the momentum transverse to the beam axis.
%At the second trigger level, the Silicon Vertex Tracker (SVT)~\cite{Ashmanskas:1999ze}
%associates axial strip clusters from the four inner SVX\,II layers with XFT track information.
%The SVT measures the distance of closest approach of a track relative to the
%beam axis (impact parameter or $d_0$) with a resolution of 50\,$\mu$m,
%which includes a contribution of 30\,$\mu$m from the beam spot transverse size.
%Hadronic decays of heavy flavor 
%Events containing hadronic decays of heavy flavor hadrons 
%are selected by requiring two tracks with 
%$120\,\mu\mbox{m}\leq d_0 \leq1\,\mbox{mm}$ each.
%At the third trigger level, a farm of computers performs complete event reconstruction online;
%the opening angle $\delta\varphi$ between the two trigger tracks is required to be
%between 2$^\circ$ and 90$^\circ$,
%and the intersection point in the $r$-$\varphi$ plane
%projected along the net momentum vector of the two tracks
%must be more than 200\,$\mu$m from the beamline.

%%%%%%%%%%%%%%%%%%%%%
% Offline selection and reconstruction
%%%%%%%%%%%%%%%%%%%%
\begin{figure}[t]
\centering
\includegraphics[width=0.45\textwidth]{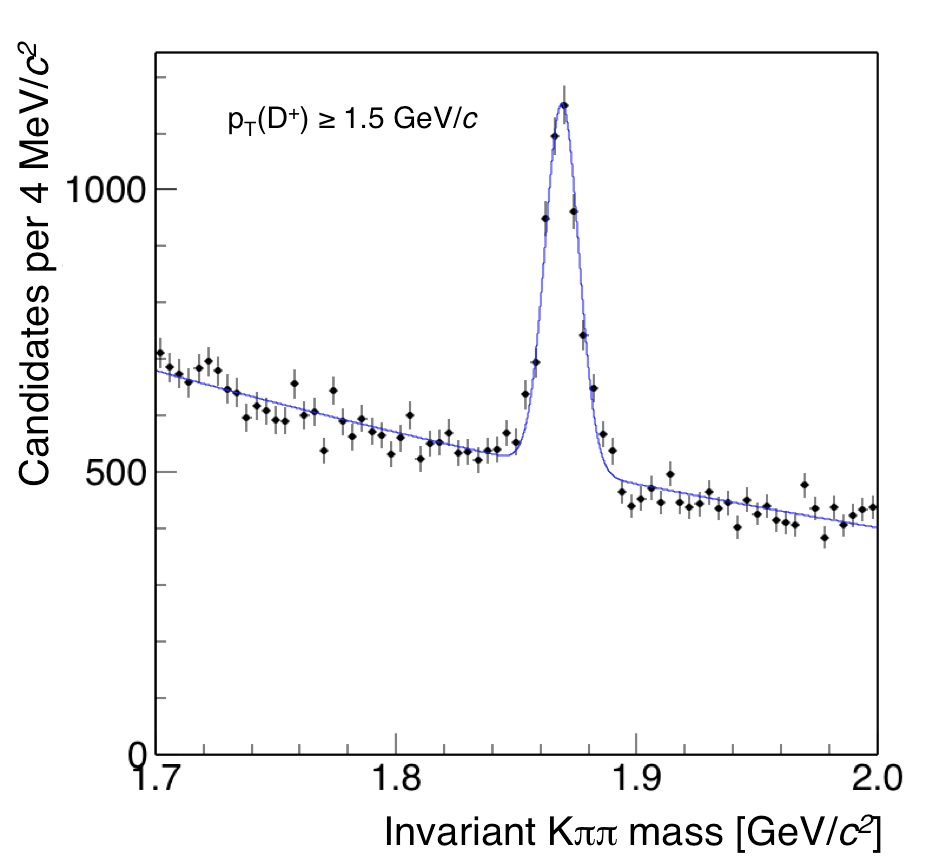}
\caption{Distribution of $K^-\pi^+\pi^+$ mass for the whole sample with fit overlaid.}.\label{fig:sample}
\end{figure}
The offline reconstruction of  $D^+ \to K^-\pi^+\pi^+$ candidates is based solely on tracking information without using particle identification, the same-charge particles being assigned the pion mass. Three good-quality tracks, associated with drift-chamber and silicon-detector information and consistent with a $K^-\pi^+\pi^+$ decay, are combined in a kinematic fit to a common decay vertex to form a $D^+$ signal candidate.  Additional selection criteria are applied on the vertex-fit quality; the minimum azimuthal separation of any pair of signal tracks; the product of their impact parameters; and the minimum value of $D^+$ transverse decay-length projected onto the direction of its $p_T$, $L_{xy}$. These criteria are fully efficient for signal and reduce backgrounds from combinations of random charged particles (combinatorics). No events are observed with more than one reconstructed candidate. We further improve the signal-to-background ratio by optimizing the selection, separately for events restricted to each of the five $D^+$ candidate $p_T$ bins, 1.5--2.5, 2.5--3.5, 3.5--4.5, 4.5--6.5, and 6.5--14.5 GeV/$c$. First, we apply an upper threshold of 100~$\mu$m on the impact parameter of the  $D^+$ candidates. This suppresses secondary $D^+$ candidates, which are less likely to point back to the $p\bar{p}$ vertex because of the combined effect of the long lifetime of $b$ hadrons and the energy released in their decay. This requirement is applied only for the optimization (see below), but is lifted in further analysis, where a fit of the $D^+$ impact-parameter distribution separates statistically the signal of prompt $D^+$ candidates from the secondaries. Then we divide the sample randomly into two subsamples. In each, we conduct an independent optimization by maximizing the quantity $S/\sqrt{S+B}$ over 1000 possible configurations of requirements on the minimum $p_T$ ($p_{T,\rm{min}}$) of any two final-state particles, minimum $L_{xy}$, and maximum value of the vertex-fit $\chi^2$. The signal (background) yields $S$ ($B$) are estimated from fits of the $K^-\pi^+\pi^+$ mass distributions with a Gaussian model for the signal and a smooth empirical function for the background. Finally, the optimal configuration resulting from each subsample is applied on the complementary subsample. 
Use of a data-driven optimization avoids the modeling uncertainties of simulation-driven methods. Biases due to statistical fluctuations are avoided by applying selection criteria identified on one subsample to the other half of the sample. The optimized criteria vary in the ranges $p_{T,\rm{min}}> 0.6-1.1$ GeV/$c$, $L_{xy} > 600-700~\mu$m, and $\chi^2< 2-7$, depending on subsample and $p_T$ bin.
The $K^-\pi^+\pi^+$ mass distribution of the resulting sample, summed over the full $p_T$ range, is shown in Fig.~\ref{fig:sample}. A prominent narrow peak of approximately 3400 $D^+ \to K^-\pi^+\pi^+$ decays, comprising both prompt signal and secondary charm candidates, overlaps a smooth background dominated by combinatorics.
% In the $K^+K^-$ sample, an additional background is contributed by misreconstructed multibody charm-meson decays, dominated by $D^0 \to h^-\pi^+ \pi^0$ and the  $D^0 \to h^- \ell^+ \nu_{\ell}$ contributions, where $\ell$ is a muon or an electron.\par
%$D^0$ candidates within three standard deviations of the $D^0$ mass are combined with a third track with $p_T\geq0.5\,$GeV/$c$ to form  $D^{*+}\to D^0 \pi^+$ candidates.
%The three-body decays of the $D^+$ and $D^+_s$ are reconstructed by combining a trigger pair with a third track having axial hits 
%in at least three out of five SVX\,II layers and performing a vertex fit based on axial track information only.
%For $D^+_s$ reconstruction, we specifically require the $K^-\pi^+$ pair to satisfy the trigger requirements,
%since the typical opening angle between two kaons from $\phi$ decay is close to the $\delta\varphi\geq2^\circ$ trigger requirement.
%Each $\phi$ candidate is required to have a mass within $\pm$20\,MeV of the world average $\phi$ mass~\cite{Hagiwara:fs}.
%The $D$ meson candidates are binned in $p_T$ as indicated in Table~\ref{table_Dx_D_bincorr}.%%The signals summed over all $p_T$ bins are shown in Fig.~\ref{fig:signal}.
\begin{figure*}[t]
\centering
\begin{overpic}[width=0.7\textwidth]{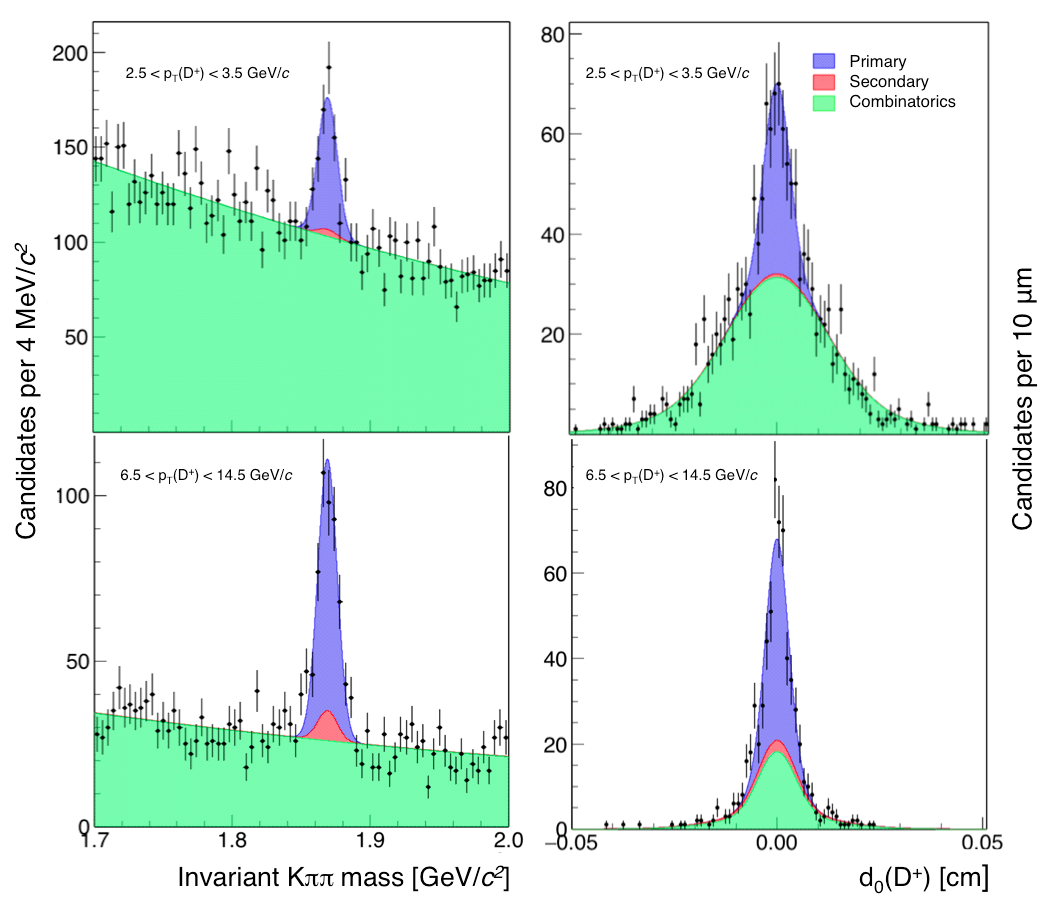}
\put(44,80){(a)}
\put(89,80){(b)}
\put(44,40){(c)}
\put(89,40){(d)}
\end{overpic}
\caption{Distributions of (a) $K^-\pi^+\pi^+$ mass  for candidates with $2.5<p_T<3.5$ GeV/$c$ and (b) $D^+$ impact parameter for those candidates, further restricted to have $K^-\pi^+\pi^+$ mass within three standard deviations from the peak value. Fits are overlaid. Panels (c) and (d) show the same distributions for candidates  with $6.5 < p_T< 14.5$ GeV/$c$.}\label{fig:MassIP}
\end{figure*}
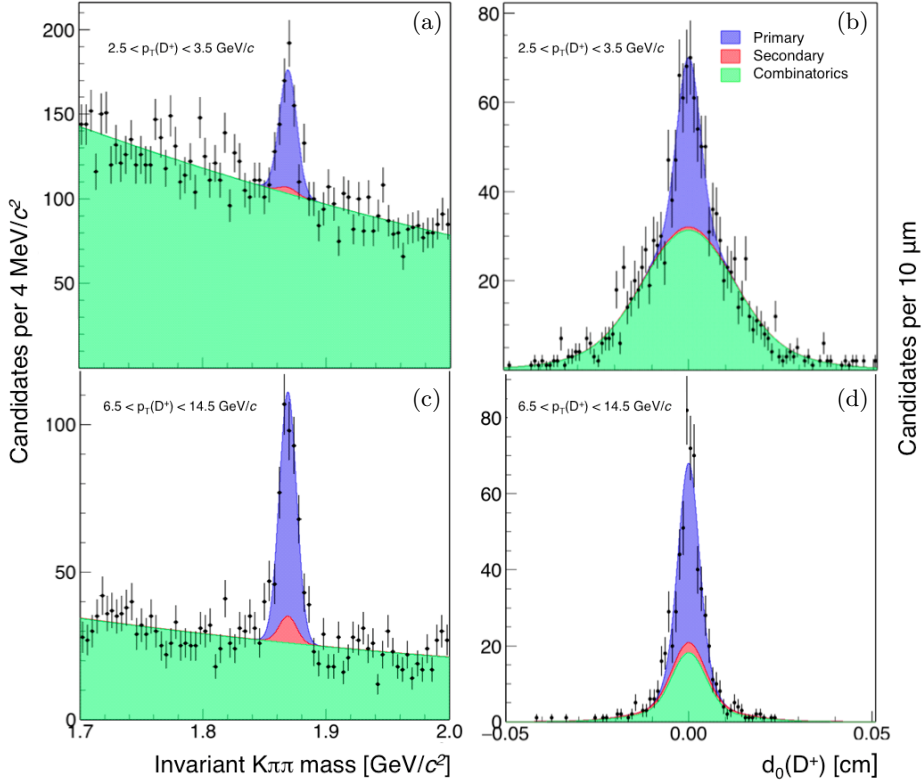
In each $p_T$ bin, we determine the yield of prompt $D^+$ decays using a simultaneous maximum-likelihood fit to the unbinned distributions of $K^-\pi^+\pi^+$  mass, to separate $D^+$ decays from combinatorics, and $D^+$ impact parameter, to separate prompt from secondary $D^+$ decays. The fit model is a linear combination of probability density functions (pdf) for prompt $D^+$ signal, secondary $D^+$, and combinatorial background, each consisting of the product of mass and impact-parameter pdfs. In the mass pdf, prompt and secondary components are modeled jointly with a Gaussian function determined from simulation; the background pdf is a second-order polynomial function derived empirically from regions with $D^+$ mass in 1.7--1.8 or 1.9--2.0~GeV/$c^2$ (sidebands). In the impact-parameter pdf, the prompt (secondary) component is modeled with the sum of three narrow (broad) Gaussian distributions determined using simulation whereas the background is modeled with a combination of Gaussian shapes that empirically reproduce the impact-parameter distribution of sideband events. The only free parameters in the fit are the numbers of prompt $D^+$ (signal) decays and secondary $D^+$ decays. Tests on simplified simulated experiments show that the fit estimates are unbiased and have proper Gaussian uncertainties. Figure~\ref{fig:MassIP} shows examples of fits in two $p_T$ bins, $2.5<p_T<3.5$ GeV/$c$ and $6.5 < p_T< 14.5$ GeV/$c$.
%Both pdf's are modelel using simulation,  invariant mass distribution, with a linear function for the combinatoric background, a narrow Gaussian for the $D^0$ signal,
%and a wide Gaussian with the same normalization describing $D^0\to K^-\pi^+$ with the wrong mass assignment. We determine the shape of the mass distribution resulting from the wrong mass  assignment using $D^0$ from $D^{*+}$ decay where the charge of the low-momentum pion determines the mass assignment of the $D^0$ decay products. 
%The $D^{*+}$ yield is extracted from the distribution of  $\Delta m=m(K^-\pi^+\pi^+)-m(K^-\pi^+)$,  the mass difference between  the $K^-\pi^+\pi^+$ and the $K^-\pi^+$ combination. 
%The signal is modeled with two Gaussians with equal means, and the background is characterized as  $a \sqrt{\Delta m-m_\pi}\exp(b(\Delta m-m_\pi))$, where $a$ and $b$ are free parameters in the fit. The $D^+$ signal is described with two Gaussians and the background is described with a linear function. In the $\phi \pi^+$ mode, we model the invariant mass distribution as a linear background with two Gaussians, one corresponding to the $D^+$ and one to the $D_s^+$.
%We find $36804\pm409$ $D^0\to K^-\pi^+$, $5515\pm85$ $D^{*+}\to D^0 \pi^+$, $28361\pm294$ $D^+\to K^-\pi^+\pi^+$, and $851\pm43$ $D_s^+ \to \phi \pi^+$, where the uncertainties quoted are statistical only.
%The observed fraction of secondary signal decays is typically 10\% between 0\% and 40\%, depending on $p_T$ with typical uncertainties of 0.0\%--0.0\%.
A total signal of approximately 2950 prompt $D^+$ decays is obtained. The observed fraction of secondary decays is typically 15\% of the total $D^+$ yield, but ranges between 0\% and 40\% with large uncertainties, depending on $p_T$. We vary the parameters and assumptions of the signal and background models and attribute systematic uncertainties on prompt signal yields in the range 0.9\%--1.5\%, depending on the $p_T$ of the candidate. \par
We factorize the reconstruction efficiency $\epsilon_i$,  relative to the $i$th $p_T$-bin, into the product of trigger efficiency, offline efficiency for reconstructing three tracks that meet the quality and fiducial requirements in the drift chamber, offline efficiency for assigning the information from the silicon detector to these tracks, and the efficiency of the offline selection requirements. The zero-bias trigger efficiency is 100\% by construction. The minimum-bias trigger efficiency is determined to be $(98.8^{+0.2}_{-0.4})$\% from the ratio of $D^+$ signal yields observed in zero-bias events that meet, or fail, the minimum-bias requirements. All offline efficiencies are known to be reproduced accurately by the simulation~\cite{Luigi} except for the term associated with the silicon detector. We therefore use efficiencies derived from simulation as inputs for the measurement and use control samples of data to obtain systematic uncertainties that cover potential data-simulation discrepancies in the silicon-related efficiency. Offline efficiencies ranging from 0.27\% to 7.5\% are determined from simulated events containing $D^+ \to K^-\pi^+\pi^+$ decays, in which distributions are weighted so that the multiplicity of prompt vertices reproduces the distribution observed in data. Control samples of muons from $J/\psi \to \mu^+\mu^-$ decays and low-momentum pions from $D^{*+} \to D^0(\to K^-\pi^+)\pi^+$ decays, in which only drift-chamber information is used to select and reconstruct the charged particle, are used to determine silicon efficiencies as functions of charged-particle $p_T$ and data-taking time from the fraction of charged particles that also meet the silicon requirements. The results are compared with silicon efficiencies determined in simulation, and the maximum observed deviation, 3.7\%, is used as the systematic uncertainty on the per-track efficiency, resulting in an 11.5\% uncertainty common to all $D^+$ transverse momentum bins. This is the largest systematic uncertainty. Additional systematic uncertainties associated with imperfect descriptions of multitrack efficiency correlations, ionization energy loss,
and hadronic interactions in the inner tracker material are negligible. Repeating the measurement on independent subsamples of data split according to data-taking time and $D$ candidate charge shows no evidence of residual biases.\par
\begin{figure}[t]
\centering
\includegraphics[width=0.4\textwidth]{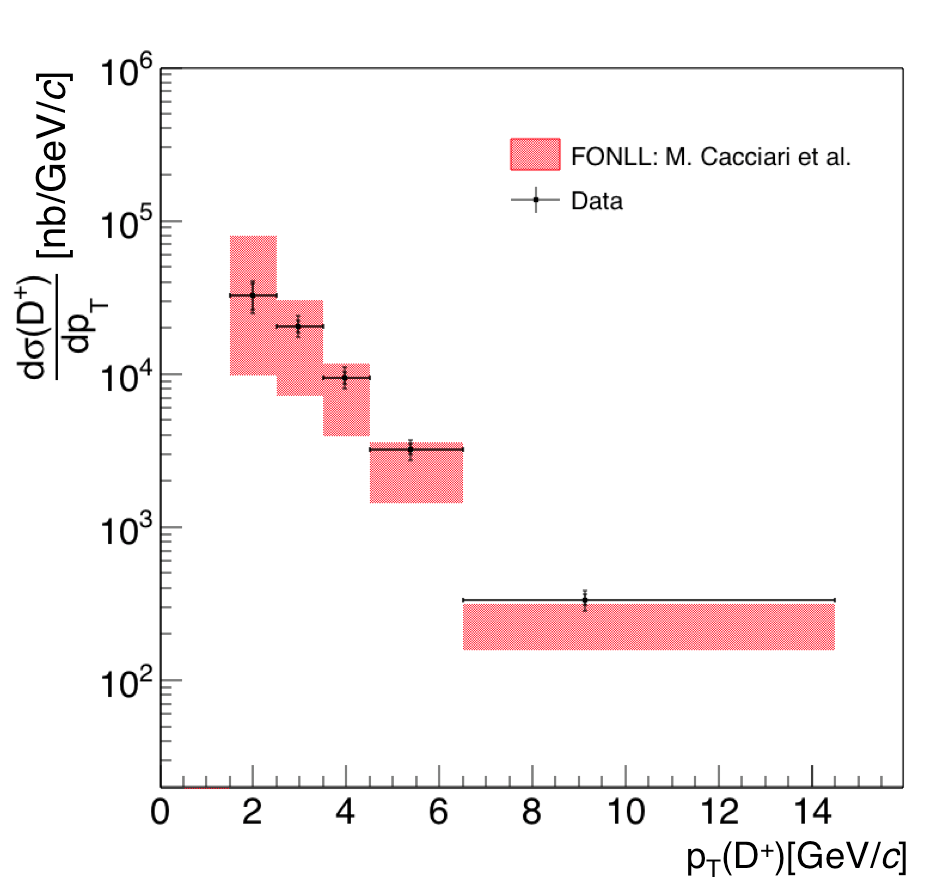}
\caption{Differential cross section as a function of $p_T$ for prompt $D^+$ mesons with $p_T> 1.5$ GeV/$c$, compared with predictions from Ref.~\cite{FONLL}.  
In each bin, the data point is displayed at the $p_T$ value at which the differential cross section equals its average over that $p_T$ bin, as determined using predictions from Ref.~\cite{FONLL}. These ``effective $p_T$" values are also listed in Table~\ref{tab:results}.}\label{fig:results}
\end{figure}

The measured differential cross sections, averaged over each $p_T$ bin and integrated over the rapidity range $|y|<1$, are shown in Table~\ref{tab:results} and displayed in Fig.~\ref{fig:results}. The observed cross sections are compatible with those predicted in recent calculations~\cite{FONLL} and with those determined in early Run II using an independent data set~\cite{Acosta:2003ax}. The total cross section for the production of $D^+$ mesons in the kinematic range $1.5 < p_T < 14.5~\mbox{GeV/}c$ and  $|y|<1$, obtained by summing over all $p_T$ bins, is  $71.9 \pm 6.8 \pm 9.3~\mu$b, where the first contribution to the uncertainty is statistical and the second systematic. 

%\begin{table}[!h]
%\begin{tabular}{lcc}
%\hline\hline
% $p_T$ range & Central $p_T$ &  $d\sigma(D^{+}, |y|<1)/d p_T$ \\
% $[\mbox{GeV}/c]$ & $[\mbox{GeV}/c]$  &   $[\mu\mbox{b/GeV/}c]$           \\
% \hline
% $1.5-2.5$      & $2.0$      & 33.9$\pm 6.8\pm 4.4$ 	\\
% $2.5-3.5$      & $3.0$      & $21.4\pm 1.9\pm 2.8$    	\\
% $3.5-4.5$      & $4.0$      & $9.82\pm 0.88\pm 1.3$  	\\
% $4.5-6.5$      & $5.5$      & $3.25\pm 0.27\pm 0.42$  	\\
% $6.5-14.5$    & $10.5$    & $0.348\pm 0.029\pm 0.045$ 	\\
% \hline\hline
%\end{tabular}

\begin{table}[!h]
\begin{tabular}{lccc}
\hline\hline
 $p_T$ range & Eff.\ $p_T$ &$d\sigma(D^{+}, |y|<1)/d p_T$ & $\sigma_i(D^{+}, |y|<1)$\\
 $(\mbox{GeV}/c)$  & $(\mbox{GeV}/c)$ &$(\mu\mbox{b/GeV/}c)$     & ($\mu$b)	     	\\
 \hline
 $1.5-2.5$	&	2.04          & $32.7\pm 6.5\pm 4.2$ 	& $32.7 \pm 6.5\pm 4.2$	\\
 $2.5-3.5$	&	2.98          & $20.6\pm 1.8\pm 2.7$   &	 $20.6\pm 1.8\pm 2.7$  	\\
 $3.5-4.5$&	3.97          & $9.5\pm 0.8\pm 1.2$ &	 $9.5\pm 0.8\pm 1.2$ 	\\
 $4.5-6.5$	&	5.38          & $3.2\pm 0.3\pm 0.4$  &	 $6.5\pm 0.5\pm 0.8$	\\
% $6.5-14.5$       & $0.336\pm 0.027\pm 0.044$ 	& $2.688\pm 0.216\pm 0.352$	\\
 $6.5-14.5$&	9.19		& $0.34\pm 0.03\pm 0.04$ 	& $2.69\pm 0.22\pm 0.35$	\\
 \hline\hline
\end{tabular}
\caption{Prompt $D^+$-meson cross-section results. All cross-section values are integrated over the range $|y|<1$. The second column (``effective $p_T$") lists the $p_T$ values at which the differential cross section equals its average over that $p_T$ bin, as determined using Ref.~\cite{FONLL}. Values in the third (fourth) column are averaged (integrated) over each $p_T$ bin. The first contribution to the uncertainties is statistical, the second systematic.
%The products of the branching fractions~\cite{Hagiwara:fs} used are 
%$(3.81\pm0.09)\%$, $(2.57\pm0.06)\%$, $(9.1\pm0.6)\%$ and $(1.8\pm0.5)\%$ 
%for $D^0$, $D^{*+}$, $D^+$ and $D_s^+$, respectively.
}
\label{tab:results}
\end{table}
%To calculate the differential cross sections, we divide $\sigma_i$ by the width of the $p_T$ bin.
%Since we report $d\sigma/dp_T$ at the center of each $p_T$ bin,
%we apply a correction to account for the non-linear shape of the cross section, %using the $p_T$ reweighted MC to obtain the shape of the cross-section inside each $p_T$ bin.
%The results are %listed in Table~\ref{table_Dx_D_bincorr}.
%The uncertainties on the calculated cross sections are evaluated by varying 
%independently the renormalization and factorization scales between 0.5 and 2 times the default scale.
%Ref.~\cite{Cacciari:2003zu} uses a default scale of $\sqrt{m^2_c+p^2_T}$,
%IJK 04 Oct 2003 - added value of c quark mass used in calculations
%where $m_c = 1.5$\,GeV/$c^2$ is the $c$ quark mass,
%while Ref.~\cite{kniehl} uses a default scale of $2\sqrt{m^2_c+p^2_T}$.
%Contributions from other sources, such as the charm quark mass, the value of the strong coupling constant
%and the fragmentation functions, were reported to be smaller and are not taken into account.
\par In summary, we report on a measurement of the prompt $D^{+}$-meson production cross-section, as a function of transverse momentum, in proton-antiproton collisions at $\sqrt{s} = 1.96$ TeV, using the full data set collected by the CDF experiment in Tevatron Run II, and corresponding to 10 fb$^{-1}$ of integrated luminosity. We use prompt $D^{+} \to K^-\pi^+\pi^+$ decays with transverse-momenta down to 1.5 GeV/$c$ fully reconstructed in the central rapidity region $|y|<1$. The differential cross section is averaged in each $p_T$ bin and integrated over the $D^+$ rapidity interval $|y|<1$. The total cross section is $\sigma(D^+, 1.5 < p_T < 14.5~\mbox{GeV/}c, |y|<1)= 71.9 \pm 6.8 (\mbox{stat}) \pm 9.3 (\mbox{syst})~\mu$b. The results are unique in that they probe strong-interaction dynamics in a low-$p_T$ regime unexplored in charm-meson production from proton-antiproton collisions. At higher transverse momentum, where previous measurement are available, the current measurements agree with earlier results~\cite{Acosta:2003ax}. While the individual measurement points lie within the band of theoretical uncertainty, the experimental spectrum is systematically shifted to high $p_T$-values as compared with theory. Hence, these results may help to further refine the shape of the theoretical cross section as a function of transverse momentum. The results may also be helpful for understanding backgrounds in astrophysical neutrino experiments at a few PeV, where most background sources are suppressed and the contributions from charm hadrons produced in the interaction of cosmic rays and atmospheric nuclei could be important.

\input{acknowledgements.tex}

\end{document}

%% file: author.tex
% Last update: $Date: 2016/10/26 16:37:55 $
\affiliation{Institute of Physics, Academia Sinica, Taipei, Taiwan 11529, Republic of China}
\affiliation{Argonne National Laboratory, Argonne, Illinois 60439, USA}
\affiliation{University of Athens, 157 71 Athens, Greece}
\affiliation{Institut de Fisica d'Altes Energies, ICREA, Universitat Autonoma de Barcelona, E-08193, Bellaterra (Barcelona), Spain}
\affiliation{Baylor University, Waco, Texas 76798, USA}
\affiliation{Istituto Nazionale di Fisica Nucleare Bologna, \ensuremath{^{kk}}University of Bologna, I-40127 Bologna, Italy}
\affiliation{University of California, Davis, Davis, California 95616, USA}
\affiliation{University of California, Los Angeles, Los Angeles, California 90024, USA}
\affiliation{Instituto de Fisica de Cantabria, CSIC-University of Cantabria, 39005 Santander, Spain}
\affiliation{Carnegie Mellon University, Pittsburgh, Pennsylvania 15213, USA}
\affiliation{Enrico Fermi Institute, University of Chicago, Chicago, Illinois 60637, USA}
\affiliation{Comenius University, 842 48 Bratislava, Slovakia; Institute of Experimental Physics, 040 01 Kosice, Slovakia}
\affiliation{Joint Institute for Nuclear Research, RU-141980 Dubna, Russia}
\affiliation{Duke University, Durham, North Carolina 27708, USA}
\affiliation{Fermi National Accelerator Laboratory, Batavia, Illinois 60510, USA}
\affiliation{University of Florida, Gainesville, Florida 32611, USA}
\affiliation{Laboratori Nazionali di Frascati, Istituto Nazionale di Fisica Nucleare, I-00044 Frascati, Italy}
\affiliation{University of Geneva, CH-1211 Geneva 4, Switzerland}
\affiliation{Glasgow University, Glasgow G12 8QQ, United Kingdom}
\affiliation{Harvard University, Cambridge, Massachusetts 02138, USA}
\affiliation{Division of High Energy Physics, Department of Physics, University of Helsinki, FIN-00014, Helsinki, Finland; Helsinki Institute of Physics, FIN-00014, Helsinki, Finland}
\affiliation{University of Illinois, Urbana, Illinois 61801, USA}
\affiliation{The Johns Hopkins University, Baltimore, Maryland 21218, USA}
\affiliation{Institut f\"{u}r Experimentelle Kernphysik, Karlsruhe Institute of Technology, D-76131 Karlsruhe, Germany}
\affiliation{Center for High Energy Physics: Kyungpook National University, Daegu 702-701, Korea; Seoul National University, Seoul 151-742, Korea; Sungkyunkwan University, Suwon 440-746, Korea; Korea Institute of Science and Technology Information, Daejeon 305-806, Korea; Chonnam National University, Gwangju 500-757, Korea; Chonbuk National University, Jeonju 561-756, Korea; Ewha Womans University, Seoul, 120-750, Korea}
\affiliation{Ernest Orlando Lawrence Berkeley National Laboratory, Berkeley, California 94720, USA}
\affiliation{University of Liverpool, Liverpool L69 7ZE, United Kingdom}
\affiliation{University College London, London WC1E 6BT, United Kingdom}
\affiliation{Centro de Investigaciones Energeticas Medioambientales y Tecnologicas, E-28040 Madrid, Spain}
\affiliation{Massachusetts Institute of Technology, Cambridge, Massachusetts 02139, USA}
\affiliation{University of Michigan, Ann Arbor, Michigan 48109, USA}
\affiliation{Michigan State University, East Lansing, Michigan 48824, USA}
\affiliation{Institution for Theoretical and Experimental Physics, ITEP, Moscow 117259, Russia}
\affiliation{University of New Mexico, Albuquerque, New Mexico 87131, USA}
\affiliation{The Ohio State University, Columbus, Ohio 43210, USA}
\affiliation{Okayama University, Okayama 700-8530, Japan}
\affiliation{Osaka City University, Osaka 558-8585, Japan}
\affiliation{University of Oxford, Oxford OX1 3RH, United Kingdom}
\affiliation{Istituto Nazionale di Fisica Nucleare, Sezione di Padova, \ensuremath{^{ll}}University of Padova, I-35131 Padova, Italy}
\affiliation{University of Pennsylvania, Philadelphia, Pennsylvania 19104, USA}
\affiliation{Istituto Nazionale di Fisica Nucleare Pisa, \ensuremath{^{mm}}University of Pisa, \ensuremath{^{nn}}University of Siena, \ensuremath{^{oo}}Scuola Normale Superiore, I-56127 Pisa, Italy, \ensuremath{^{pp}}INFN Pavia, I-27100 Pavia, Italy, \ensuremath{^{qq}}University of Pavia, I-27100 Pavia, Italy}
\affiliation{University of Pittsburgh, Pittsburgh, Pennsylvania 15260, USA}
\affiliation{Purdue University, West Lafayette, Indiana 47907, USA}
\affiliation{University of Rochester, Rochester, New York 14627, USA}
\affiliation{The Rockefeller University, New York, New York 10065, USA}
\affiliation{Istituto Nazionale di Fisica Nucleare, Sezione di Roma 1, \ensuremath{^{rr}}Sapienza Universit\`{a} di Roma, I-00185 Roma, Italy}
\affiliation{Mitchell Institute for Fundamental Physics and Astronomy, Texas A\&M University, College Station, Texas 77843, USA}
\affiliation{Istituto Nazionale di Fisica Nucleare Trieste, \ensuremath{^{ss}}Gruppo Collegato di Udine, \ensuremath{^{tt}}University of Udine, I-33100 Udine, Italy, \ensuremath{^{uu}}University of Trieste, I-34127 Trieste, Italy}
\affiliation{University of Tsukuba, Tsukuba, Ibaraki 305, Japan}
\affiliation{Tufts University, Medford, Massachusetts 02155, USA}
\affiliation{Waseda University, Tokyo 169, Japan}
\affiliation{Wayne State University, Detroit, Michigan 48201, USA}
\affiliation{University of Wisconsin-Madison, Madison, Wisconsin 53706, USA}
\affiliation{Yale University, New Haven, Connecticut 06520, USA}

\author{T.~Aaltonen}
\affiliation{Division of High Energy Physics, Department of Physics, University of Helsinki, FIN-00014, Helsinki, Finland; Helsinki Institute of Physics, FIN-00014, Helsinki, Finland}
\author{S.~Amerio\ensuremath{^{ll}}}
\affiliation{Istituto Nazionale di Fisica Nucleare, Sezione di Padova, \ensuremath{^{ll}}University of Padova, I-35131 Padova, Italy}
\author{D.~Amidei}
\affiliation{University of Michigan, Ann Arbor, Michigan 48109, USA}
\author{A.~Anastassov\ensuremath{^{w}}}
\affiliation{Fermi National Accelerator Laboratory, Batavia, Illinois 60510, USA}
\author{A.~Annovi}
\affiliation{Laboratori Nazionali di Frascati, Istituto Nazionale di Fisica Nucleare, I-00044 Frascati, Italy}
\author{J.~Antos}
\affiliation{Comenius University, 842 48 Bratislava, Slovakia; Institute of Experimental Physics, 040 01 Kosice, Slovakia}
\author{G.~Apollinari}
\affiliation{Fermi National Accelerator Laboratory, Batavia, Illinois 60510, USA}
\author{J.A.~Appel}
\affiliation{Fermi National Accelerator Laboratory, Batavia, Illinois 60510, USA}
\author{T.~Arisawa}
\affiliation{Waseda University, Tokyo 169, Japan}
\author{A.~Artikov}
\affiliation{Joint Institute for Nuclear Research, RU-141980 Dubna, Russia}
% ...

\collaboration{CDF Collaboration}
\altaffiliation[With visitors from]{
\ensuremath{^{a}}University of British Columbia, Vancouver, BC V6T 1Z1, Canada,
\ensuremath{^{b}}Istituto Nazionale di Fisica Nucleare, Sezione di Cagliari, 09042 Monserrato (Cagliari), Italy,
\ensuremath{^{c}}University of California Irvine, Irvine, CA 92697, USA,
\ensuremath{^{d}}Institute of Physics, Academy of Sciences of the Czech Republic, 182~21, Czech Republic,
\ensuremath{^{e}}CERN, CH-1211 Geneva, Switzerland,
\ensuremath{^{f}}Cornell University, Ithaca, NY 14853, USA,
\ensuremath{^{g}}University of Cyprus, Nicosia CY-1678, Cyprus,
\ensuremath{^{h}}Office of Science, U.S. Department of Energy, Washington, DC 20585, USA,
\ensuremath{^{i}}University College Dublin, Dublin 4, Ireland,
\ensuremath{^{j}}ETH, 8092 Z\"{u}rich, Switzerland,
\ensuremath{^{k}}University of Fukui, Fukui City, Fukui Prefecture, Japan 910-0017,
\ensuremath{^{l}}Universidad Iberoamericana, Lomas de Santa Fe, M\'{e}xico, C.P. 01219, Distrito Federal,
\ensuremath{^{m}}University of Iowa, Iowa City, IA 52242, USA,
\ensuremath{^{n}}Kinki University, Higashi-Osaka City, Japan 577-8502,
\ensuremath{^{o}}Kansas State University, Manhattan, KS 66506, USA,
\ensuremath{^{p}}Brookhaven National Laboratory, Upton, NY 11973, USA,
\ensuremath{^{q}}Istituto Nazionale di Fisica Nucleare, Sezione di Lecce, Via Arnesano, I-73100 Lecce, Italy,
\ensuremath{^{r}}Queen Mary, University of London, London, E1 4NS, United Kingdom,
\ensuremath{^{s}}University of Melbourne, Victoria 3010, Australia,
\ensuremath{^{t}}Muons, Inc., Batavia, IL 60510, USA,
\ensuremath{^{u}}Nagasaki Institute of Applied Science, Nagasaki 851-0193, Japan,
\ensuremath{^{v}}National Research Nuclear University, Moscow 115409, Russia,
\ensuremath{^{w}}Northwestern University, Evanston, IL 60208, USA,
\ensuremath{^{x}}University of Notre Dame, Notre Dame, IN 46556, USA,
\ensuremath{^{y}}Universidad de Oviedo, E-33007 Oviedo, Spain,
\ensuremath{^{z}}CNRS-IN2P3, Paris, F-75205 France,
\ensuremath{^{aa}}Universidad Tecnica Federico Santa Maria, 110v Valparaiso, Chile,
\ensuremath{^{bb}}Sejong University, Seoul 143-747, Korea,
\ensuremath{^{cc}}The University of Jordan, Amman 11942, Jordan,
\ensuremath{^{dd}}Universite catholique de Louvain, 1348 Louvain-La-Neuve, Belgium,
\ensuremath{^{ee}}University of Z\"{u}rich, 8006 Z\"{u}rich, Switzerland,
\ensuremath{^{ff}}Massachusetts General Hospital, Boston, MA 02114 USA,
\ensuremath{^{gg}}Harvard Medical School, Boston, MA 02114 USA,
\ensuremath{^{hh}}Hampton University, Hampton, VA 23668, USA,
\ensuremath{^{ii}}Los Alamos National Laboratory, Los Alamos, NM 87544, USA,
\ensuremath{^{jj}}Universit\`{a} degli Studi di Napoli Federico II, I-80138 Napoli, Italy
}
\noaffiliation
% Last update: $Date: 2016/10/26 16:37:55 $

%% file: acknowledgements.tex
We thank the Fermilab staff and the technical staffs of the
participating institutions for their vital contributions. This work
was supported by the U.S. Department of Energy and National Science
Foundation; the Italian Istituto Nazionale di Fisica Nucleare; the
Ministry of Education, Culture, Sports, Science and Technology of
Japan; the Natural Sciences and Engineering Research Council of
Canada; the National Science Council of the Republic of China; the
Swiss National Science Foundation; the A.P. Sloan Foundation; the
Bundesministerium f\"ur Bildung und Forschung, Germany; the Korean
World Class University Program, the National Research Foundation of
Korea; the Science and Technology Facilities Council and the Royal
Society, United Kingdom; the Russian Foundation for Basic Research;
the Ministerio de Ciencia e Innovaci\'{o}n, and Programa
Consolider-Ingenio 2010, Spain; the Slovak R\&D Agency; the Academy
of Finland; the Australian Research Council (ARC); and the EU community
Marie Curie Fellowship Contract No. 302103.